\documentclass[journal=nalefd,manuscript=letter,layout=onecolumn]{achemso}

\author{Michael de Oliveira} 
\affiliation[CNST@IIT]{Center for Nano Science and Technology, Fondazione Istituto Italiano di Tecnologia, Milan, Italy}
\alsoaffiliation[Polimi]{Physics Department, Politecnico di Milano, Milan, Italy}
\altaffiliation{Contributed equally to this work}

\author{Marco Piccardo}
\affiliation[CNST@IIT]{Center for Nano Science and Technology, Fondazione Istituto Italiano di Tecnologia, Milan, Italy}
\altaffiliation{Contributed equally to this work}

\author{Sahand Eslami}
\affiliation[IIT]{Fondazione Istituto Italiano di Tecnologia, Genoa, Italy}

\author{Vincenzo Aglieri}
\affiliation[IIT]{Fondazione Istituto Italiano di Tecnologia, Genoa, Italy}

\author{Andrea Toma}
\affiliation[IIT]{Fondazione Istituto Italiano di Tecnologia, Genoa, Italy}

\author{Antonio Ambrosio}
\email{antonio.ambrosio@iit.it}
\affiliation[CNST@IIT]{Center for Nano Science and Technology, Fondazione Istituto Italiano di Tecnologia, Milan, Italy}

\title{Radially and azimuthally pure vortex beams from phase-amplitude metasurfaces}


\begin{document}


\begin{abstract}
To exploit the full potential of the transverse spatial structure of light using the Laguerre–Gaussian basis, it is necessary to control the azimuthal and radial components of the photons. Vortex phase elements are commonly used to generate these modes of light, offering precise control over the azimuthal index but neglect the radially dependent amplitude term which defines their associated corresponding transverse profile. Here we experimentally demonstrate the generation of high purity Laguerre–Gaussian beams with a single step on-axis transformation implemented with a dielectric phase-amplitude metasurface. By vectorially structuring the input beam and projecting it onto an orthogonal polarisation basis, we can sculpt any vortex beam in phase and amplitude. We characterize the azimuthal and radial purity of the generated vortex beams, reaching a purity of 98\% for a vortex beam with $\ell=50$ and $p=0$. Furthermore, we comparatively show that the purity of the generated vortex beams outperform those generated with other well-established phase-only metasurface approaches. In addition, we highlight the formation of `ghost' orbital angular momentum orders from azimuthal gratings (analogous to ghost orders in ruled gratings), which have not been widely studied to date. Our work brings higher-order vortex beams and their unlimited potential within reach of wide adoption.
\end{abstract}


\hfill \break
The ability to structure light in all its transverse degrees of freedom has led to advances in fundamental science and real-world applications, both classical and quantum~\cite{forbes2021structured}. Particularly significant and utile are vortex light beams, which carry orbital angular momentum (OAM)~\cite{shen2019optical}. Since they are associated with a vortex phase of the form $e^{i\ell\phi}$~\cite{allen1992orbital}, azimuthal phase elements have become ubiquitous for their generation, from spiral phase plates~\cite{beijersbergen1994helical} and computer-generated holograms displayed on spatial light modulators (SLM)~\cite{carpentier2008making}, to the use of geometric phase elements such as liquid crystals~\cite{marrucci2006optical,brasselet2009optical,kim2015fabrication}, or they can be created directly at the source~\cite{forbes2019structured}. Recently, a resurgence in vortex generation approaches has emerged with the advent of nano-fabrication~\cite{piccardo2020recent}, moving away from traditional bulky optical components to subwavelength structured planar devices that shape the wavefront of light in all its properties, i.e., amplitude, phase and polarization.

The ease of use and ability to control polarization and phase in metasurfaces have facilitated the realization of many well-established metasurface devices for optical vortex generation, including $q$-plates~\cite{devlin2017spin} and their notable generalization $J$-plates~\cite{devlin2017arbitrary}. However, as with other phase-only devices, they approximate a vortex beam by an incident Gaussian beam modulated by an azimuthal phase profile, and subsequently neglect the amplitude term which defines the annular intensity associated with these modes. As a consequence, the generated beam is not a solution to the paraxial wave equation and thus not an eigenmode of free-space. During its propagation, the beam unravels, leading to the excitation of many undesirable radial modes~\cite{sephton2016revealing}. An example of a resulting impure vortex mode, with its many concentric rings of intensity, is depicted in Figure S3. This phenomenon has been shown to have a deleterious effect on the detection efficiency in both classical and quantum OAM applications~\cite{nape2020enhancing}, the implication of which becomes increasingly prominent for beams carrying larger OAM.

To ensure the invariance associated with eigenmodes during their propagation, the generated vortex beams must be generalised solutions of the paraxial wave equation. Many families of transverse solutions to the paraxial wave equation exist. In cylindrical coordinates, the solutions form a complete, orthonormal basis, called Laguerre-Gaussian (LG) modes of light. These modes require a pair of independent indices to fully describe them: the azimuthal and radial indices, $\ell$ and $p$. LG beams have garnered much attention because of their circular symmetry, the infinite Hilbert space they provide and their inherent relation to the quantization of orbital angular momentum (OAM)~\cite{yao2011orbital}. They are natural modes of quadratic index media~\cite{newstein1987laguerre}, making them a viable candidate for free-space and optical fiber communication. Their transverse structure is given by~\cite{saleh2019fundamentals},
\begin{equation}
  \mbox{LG}_{\ell,p}=\sqrt{\frac{2p!}{\pi\omega_0^2\left(p+|\ell|\right)!}}\left(\frac{r\sqrt{2}}{\omega_0}\right)^{|\ell|}L^{|\ell|}_{p}\left(\frac{2r^2}{\omega_0^2}\right)e^{\frac{-r^2}{\omega_0^2}}e^{-i\ell\phi}\equiv A_{\ell,p}(r)e^{i\psi(\phi)},
\label{eqn:LG}
\end{equation}
where $\omega_0$ is the beam waist, $L^{|\ell|}_{p}(x)$ is the generalized Laguerre polynomial of argument $x$, and $r$ and $\phi$ are the radial and azimuthal coordinates. The azimuthal component, $\ell$, is related to a vortex phase profile $\psi(\phi)$, which results in a `twist' of the wavefront and quantizes the OAM of $\ell\hbar$ per photon~\cite{allen1992orbital}. On the other hand, the radial component, $p$, is related to the transverse amplitude distribution $A_{\ell,p}(r)$, which results in `ripples' in the beam intensity. Although neglected in the past, radial modes have attracted interest for their diffraction properties as a means to control light propagation in complex media~\cite{vcivzmar2012exploiting,rotter2017light}. Recent work has also explored using the radial component as an additional encoding space for quantum~\cite{karimi2014radial,karimi2014exploring} and classical information protocols~\cite{zhao2015capacity,trichili2016optical}.

\begin{figure}[t!]
    \centering
    \includegraphics[width=0.5\linewidth]{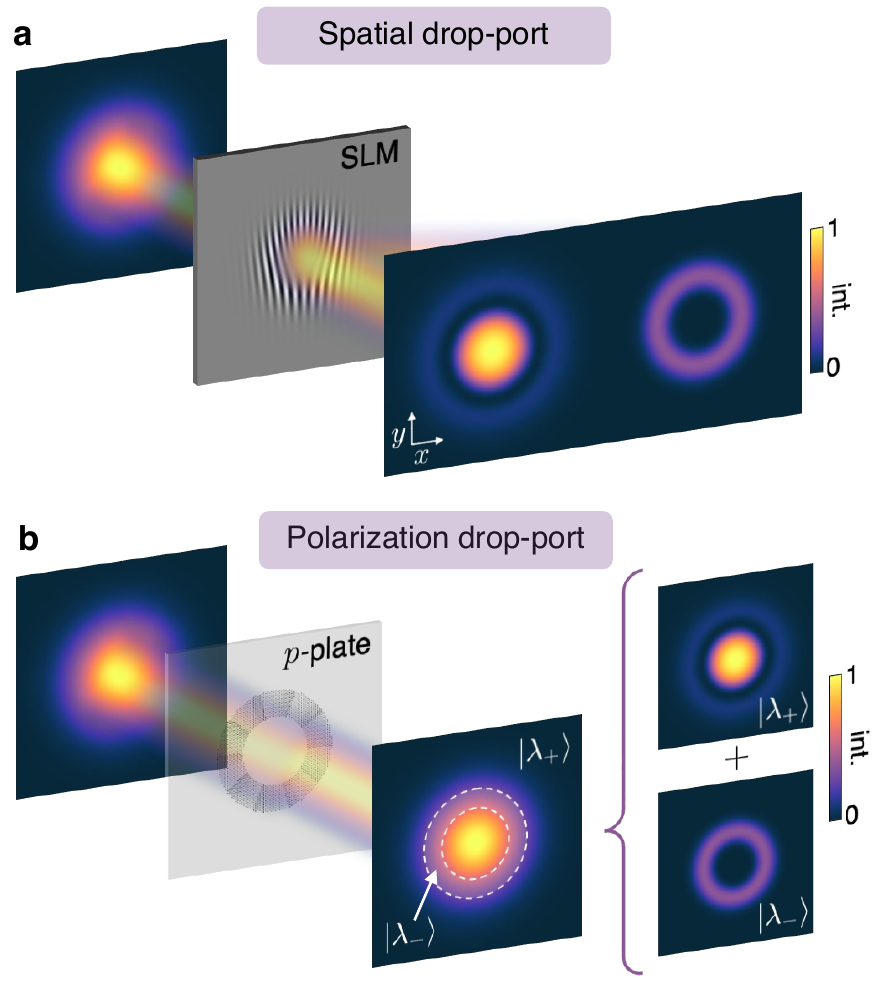}
    \caption{
    (a) A spatial drop-port, implemented via a grating, is used to carve the amplitude of the incident beam and discard the remainder in a drop-port that can be spatially filtered. 
    (b) Phase-amplitude metasurfaces allow for an alternative approach using polarization as a drop-port. By vectorially structuring the incident beam and projecting it onto an orthogonal polarization state ($\lambda_{-}$), the desired amplitude can be revealed.
    }
    \label{fig1}
\end{figure}

In order to have control over the azimuthal and radial indices, we need to be able to modulate not only the phase of the incident beam, but also its amplitude. One approach is to use an active resonator to facilitate the mode conversion necessary for generating pure OAM modes~\cite{maguid2018topologically,sroor2020high}, but this requires elaborate cavity configurations. Alternative free-space methods have also been demonstrated, which employ complex amplitude modulation using phase-only devices~\cite{bolduc2013exact}. This involves carving the amplitude of the incident beam and discarding the remainder in an unwanted drop-port. The common practice is to use a spatial drop-port as shown in Figure~\ref{fig1}a, in which a linear grating diffracts the desired spatial profile, leaving the unwanted amplitude on the main optical path which can be spatially filtered using a aperture. The drawback of this method is that it requires working off-axis with a diffraction order. Nanostructured silica glass devices have also been proposed to produce pure vortex beams~\cite{rafayelyan2017laguerre}. However, the devices are restricted to using circularly polarized incident fields, with their practical demonstration being limited to generating vortex beams with low OAM values~\cite{rafayelyan2017application,coursault2022nanostructured}. On the other hand, vortex beams of high OAM charge and purity have been generated using a two step process, which combines the high resolution of metasurfaces with complex amplitude modulation achieved using a spatial light modulator to correct for the missing amplitude term~\cite{nape2020enhancing}.

Here, using polarization as a drop-port, we experimentally demonstrate a generalized method to generating LG modes of high radial and azimuthal purity, which we achieve using a single dielectric metasurface - called the $p$-plate~\cite{piccardo2021arbitrary}. The device's ability to structure light vectorially allows an arbitrary input beam to be modulated in phase and amplitude~\cite{overvig2019dielectric,divitt2019ultrafast} (after a projection onto the orthogonal polarization state) using a single on-axis transformation, as shown in Figure~\ref{fig1}b. We refer to such a combination as a phase-amplitude metasurface~\cite{overvig2019dielectric}. The compact device has the advantage that it can be structured as an annular device without sacrificing performance, which in part relaxes the phase resolution limitation of the device near the singularity point. This additionally allows larger devices to be fabricated while reducing the write time for electron beam lithography. It follows that the total structured area of the device is inversely proportional to the OAM it imparts (see Figure S4). This results in a reduction in structured area of up to 75\% for a device designed to impart a charge of $\ell=50$ as compared to other metasurface vortex generators that typically require the full circular area to be structured. Overall, we show experimentally that our device can generate modes with well-defined radial and azimuthal purity, albeit with a simple and practical approach. 

\section{Results}

\textbf{Ripple-free vortex beams with high topological charge.} The metasurface device is designed to impart an azimuthal phase, as well as a radially-varying polarization transformation that converts part of the incident beam, with the desired amplitude distribution, to the orthogonal polarization state. The resulting beam is a vector vortex beam that carries OAM and exhibits a non-uniform polarization distribution. A polarizer, either implemented as a separate element or as a wire-grid grating integrated in the metasurface substrate~\cite{divitt2019ultrafast}, can then be used to select the orthogonal polarization and effectively filter out the excess amplitude and any unconverted light. 

\begin{figure*}[t!]
    \centering
    \includegraphics[width=1\linewidth]{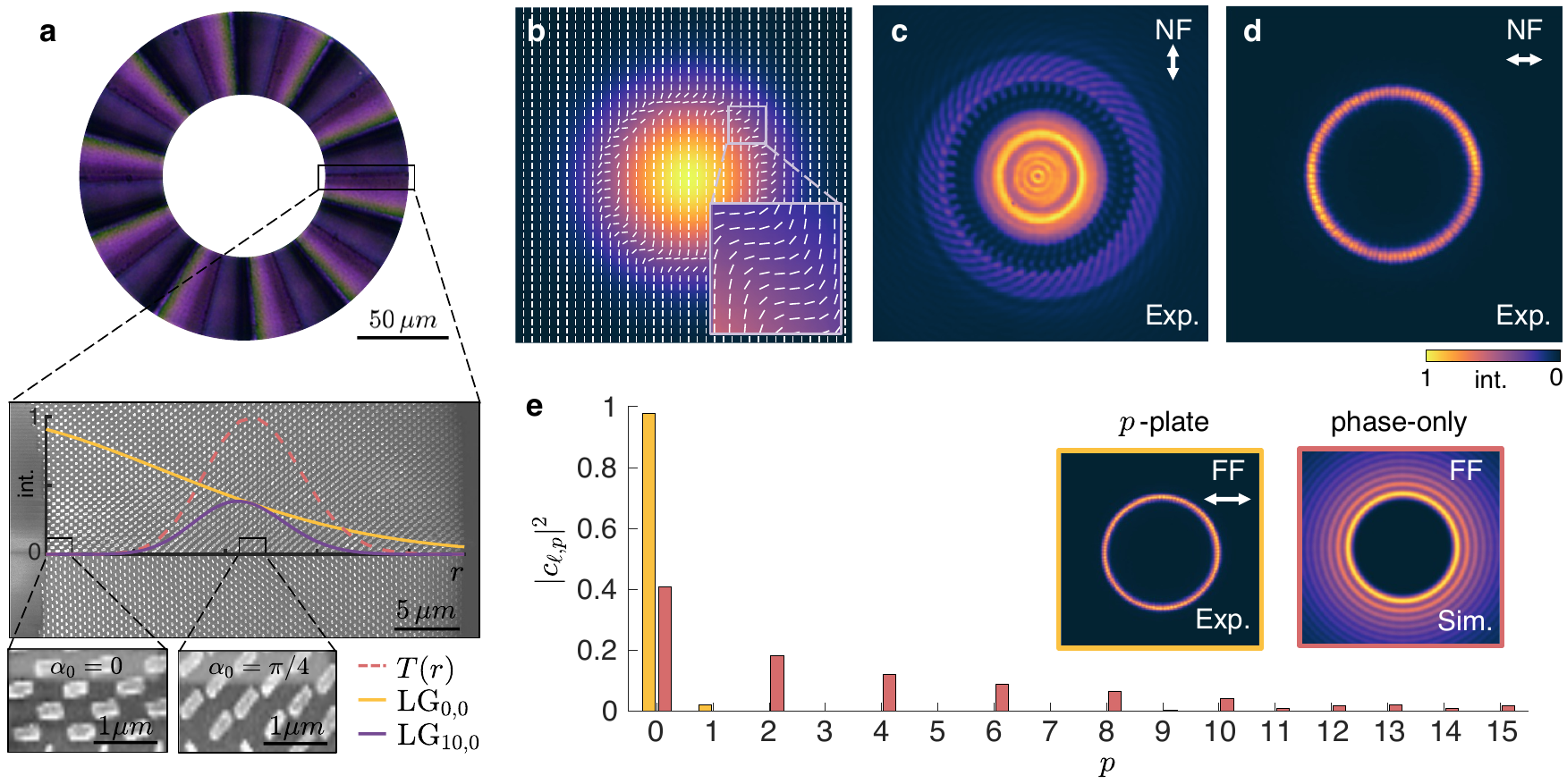}
    \caption{
    (a) Optical image of a $p$-plate metasurface device designed to generate a pure $\mbox{LG}_{10,0}$ vortex mode. The insets show scanning electron microscope (SEM) images of the device, overlaid with the transmission function $T(r)$ which is related to the rotation angle $\alpha_0$ of the pillars. This defines the vectorial polarization conversion necessary to carve the desired amplitude $\mbox{LG}_{10,0}$ from the incident Gaussian amplitude $\mbox{LG}_{0,0}$.
    (b) The simulated vectorial polarization structure of a beam generated directly at the plane of the $p$-plate device designed with $\ell=50$ and $p=0$. The corresponding experimental near-field intensity distributions of the beam projected onto the (c) vertical and (d) horizontal polarization state.
    (e) The radial $p$-mode spectrum of the vortex beam generated by the $p$-plate compared to the theoretical spectrum of an azimuthal phase-only approach. The insets show the corresponding far-field (FF) intensity distributions.
    }
    \label{fig2}
\end{figure*}

An example of a fabricated $p$-plate device used to generate a pure $\mbox{LG}_{10,0}$ vortex mode is shown in Figure~\ref{fig2}a. The metasurface device consists of many rectangular nanopillars, with the birefringence of each pillar arising from its form factor. Specifically, the desired azimuthal phase profile $\psi(\phi)$ defines one of the two dimensions of the pillar $L_x(\phi)$, while the other dimension $L_y(\phi)$ is determined by $\psi(\phi)+\Delta\phi_{0}$, where $\Delta\phi_{0}$ is a fixed parameter that corresponds to the phase retardance required to perform the desired polarization conversion to the orthogonal state ~\cite{piccardo2021arbitrary}. In addition, the rotation angles of the pillars, $\alpha_0$, are varied along the radial direction of the plate so as to embed a transmission function $T(r)$ (shown as a dashed red line in the insets of Figure~\ref{fig2}a), which upon projection of the converted polarization state sculpts the desired amplitude profile $A_{\ell,p}(r)$ (shown as a solid purple line) from the incident Gaussian amplitude (shown as a solid yellow line). Furthermore, since the desired beam has zero intensity at the center, the metasurface can be preemptively patterned to its characteristic annular shape, as the metasurface in this region would not impart a polarization conversion and would, in any case, be filtered by the polarizer. In our case the annular region is selected where the intensity transmission function is above 5\%.

The operating principle of the $p$-plate allows it to be designed to impart the required transformations on any incident wave with known polarization, phase and amplitude distribution, such as a plane or Gaussian wave. We designed and fabricated the $p$-plates to act on a linearly polarized Gaussian beam, as it is a readily available source in any laboratory. To optimize the generation efficiency of the device, we design the $p$-plates for the specific beam waist $\omega_{s}$ of the Gaussian source such that the overlap between the incident intensity profile and that of the desired intensity profile is maximized. The corresponding beam waist $\omega_{0}$ of the Gaussian embedded in the target LG mode is then given by $\omega_{0}=\omega_{s}/\sqrt{|\ell|+1}$. We note that in situations where efficiency can be neglected, this constraint can be lifted by designing the device for an incident plane wave (see Figure S6). In this way, any beam in the laboratory can be easily extended to approximate a plane wave, enabling the $p$-plate device to work with any readily available source, albeit less efficiently since the overlap of intensity profiles will be smaller and more light will be discarded in the drop-port. As a demonstration, we fabricated another $p$-plate device designed to modulate a plane wave, with the corresponding generated mode shown in Figure S6.

To establish the efficacy of the $p$-plate device in generating pure vortex modes, we experimentally demonstrate a $p$-plate that when incident with a Gaussian beam generates a $\mbox{LG}_{50,0}$ vortex beam with a charge of $\ell=50$ and no radial modes ($p=0$). A schematic of the generation and detection experimental setup is shown in Figure S7. The vectorial structure of the beam after the metasurface is shown in Figure~\ref{fig2}b. Figure~\ref{fig2}c and d show the corresponding experimental near-field intensity distributions of the generated $\mbox{LG}_{50,0}$ mode when projected onto the vertical and horizontal polarization basis, respectively. The corresponding horizontally polarized component has the characteristic annular intensity distribution of a vortex mode, while the vertically polarized component is the unwanted complementary intensity distribution from which the intensity was carved, leaving behind a dark ring. 

We studied the radial mode purity of the generated vortex by decomposing it onto the $\mbox{LG}_{50,p}$ basis, with the measured $p$-mode spectrum shown in Figure~\ref{fig2}e. The results confirm a high radial mode purity, with 97\% of the generated beam power being in the desired $p=0$ mode (see Figure S8 for the azimuthal mode decomposition). The Fourier intensity distribution of the filtered vortex beam after it is propagated to the far-field is shown in the inset. We note that no spatial filtering was employed and that the polarization conversion to the orthogonal state allowed any residual light to be filtered by the polarizer. For comparison, the theoretical $p$-mode spectrum is also shown for the case when only an azimuthal phase is used to impart such a high OAM. In this case the power in the desired $p=0$ mode drops to 40\%, with most of the beams energy being spread over higher order radial modes. Optically, this manifests as concentric rings or radial `ripples' in the beams intensity, as seen in the accompanying inset of Figure~\ref{fig2}e. 

\textbf{Comparing dielectric metasurfaces for vortex beam generation.} Having demonstrated that $p$-plate metasurfaces can generate vortex beams with high OAM charges and purity, we experimentally compare their performance to those of phase-only metasurfaces. As subjects of the latter, we consider the widely used dielectric metasurfaces, $q$-plates and $J$-plates, as vortex beam generators. The $q$-plate is a spin-orbit coupling device, which uses geometric phase to convert a circularly polarized incident Gaussian beam to the orthogonal polarization state, whilst imparting an azimuthal phase. These devices are designed using identical pillars, which effectively act as half-waveplates, and whose orientation angle varies in azimuth to confer the desired phase profile. On the other hand, the $J$-plate, a significant generalization of the $q$-plate, decouples the dimensions of the pillars, combining both propagation and geometric phase control to imbue any two orthogonal polarization states with independent arbitrary values of OAM, $\ell_1$ and $\ell_2$. Additionally, we judiciously design the $J$-plates to convert incident circular polarization to the orthogonal state with opposite handedness. Therefore, in $q$-plates and $J$-plates, the prescribed polarization conversion allows any residual Gaussian beam to be filtered. In contrast to $p$-plate devices, the nanopillars of $q$-plates and $J$-plates do not vary in dimension or orientation angle along the radial direction and as such do not apply any amplitude shaping. Consequently, for $q$-plates and $J$-plates, the beam waist, $\omega_s$, of the incident Gaussian beam determines the beam waist $\omega_0$ of the generated LG beam. To ensure a fair comparison, we fabricated all devices using the same design library and methods as before. In addition, the characterisation of all devices was performed with the same incident Gaussian beam, so that LG beams carrying the same OAM had the same beam waist, $\omega_0$. 

\begin{figure*}[t!]
    \centering
    \includegraphics[width=1\textwidth]{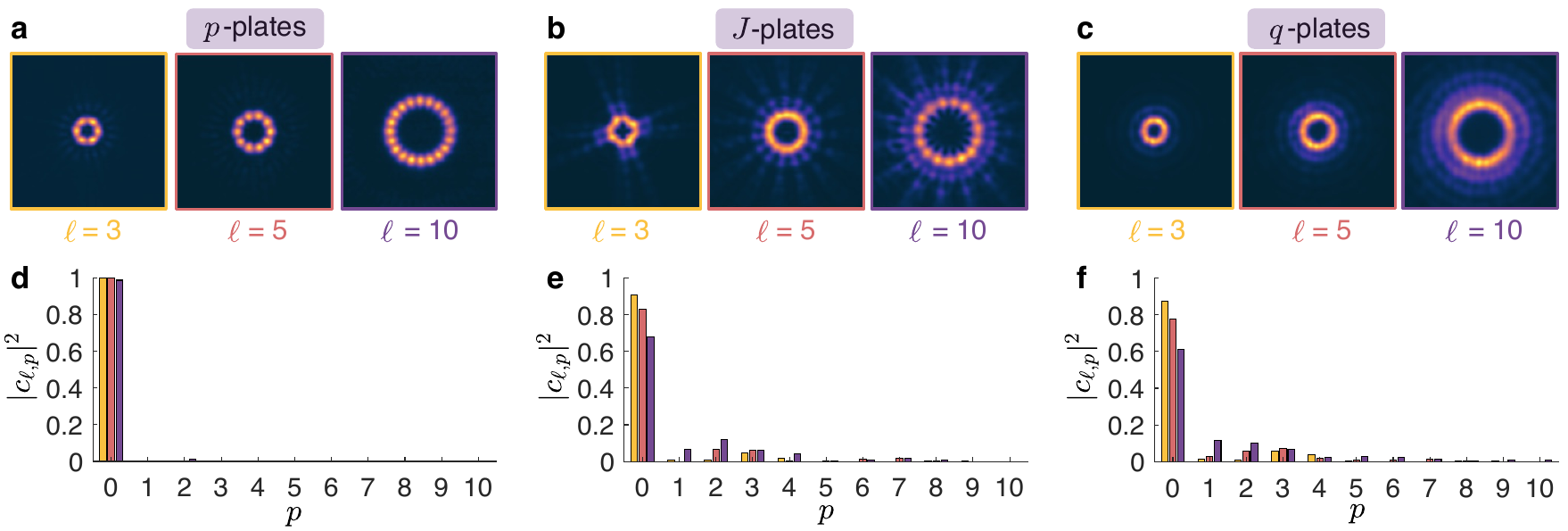}
    \caption{The experimental far-field intensity distributions of vortex beams generated using amplitude and phase control in (a) $p$-plates, and phase-only control in (b) $J$-plates and (c) $q$-plates. For each type of device, three devices were designed to impart the topological charge $\ell = 3,5,10$. 
    Corresponding radial $p$-mode spectra for: (d) $p$-plates, (e) $J$-plates and (f) $q$-plates, obtained via a modal decomposition in the $\mbox{LG}_{\ell,p}$ basis. We note that no spatial filtering was used in the generation or detection of the vortex beams.
    }
    \label{fig3}
\end{figure*}

The comparison between the three types of devices ($p$-plates, $J$-plates and $q$-plates) used to generate vortex beams of charges $\ell = 3, 5, 10$, is presented in Figure~\ref{fig3}. Visually, many concentric intensity rings can be seen in intensity distribution shown in Figure~\ref{fig3}b and c, and are more apparent for beams carrying higher OAM. These radial `ripples' are a consequence of any phase-only approach that modulates a Gaussian beam by an azimuthal phase, and are not limited to metasurfaces. In contrast, when the radial degree of freedom is controlled as a means of modulating the amplitude, a single intensity ring appears, as in the case of the $p$-plate shown in Figure~\ref{fig3}a. Figures~\ref{fig3}d through f show the corresponding $p$-mode decomposition obtained for each of the devices. For $p$-plate devices (see Figure~\ref{fig3}d), the power in the desired $\mbox{LG}_{\ell,0}$ mode is greater than 99\% for all $\ell$ modes, indicating high radial mode purity. The $p$-mode spectrum of azimuthal phase-only devices in Figure~\ref{fig3}e and f show how the mode power is spread over a superposition of higher radial modes. This effect becomes more pronounced as the OAM increases, with the power in the desired $p=0$ radial mode reaching only 61\% and 66\% for the $\ell =10$ mode generated using our $J$-plate and $q$-plate, respectively. These results show a clear advantage for using phase-amplitude modulation for vortex beam generation, particularly for beams with high OAM.

\begin{figure*}[t!]
    \centering
    \includegraphics[width=0.7\linewidth]{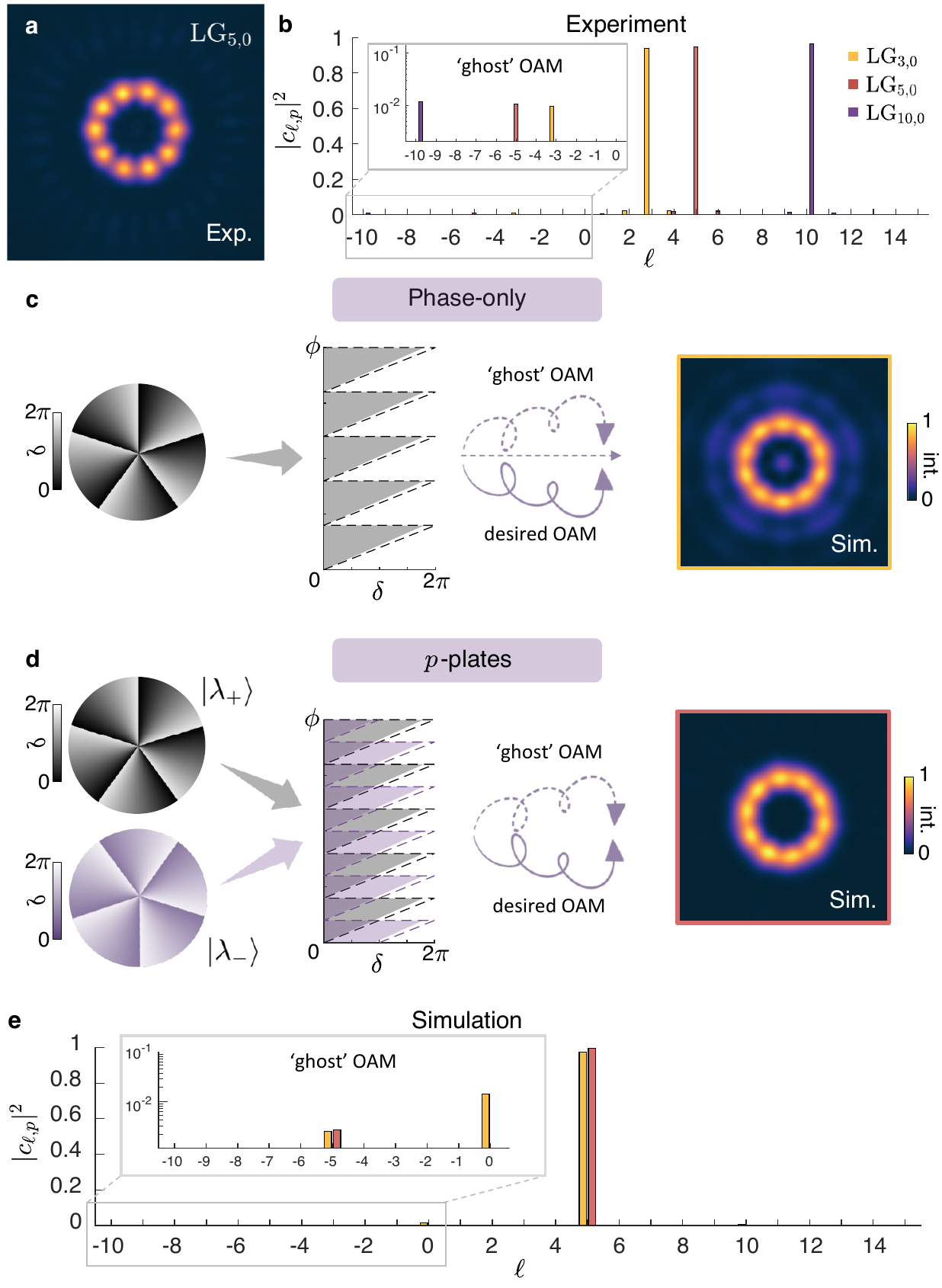}
    \caption{ 
    (a) The experimental intensity distribution of a vortex beam generated by a $p$-plate with $\ell=5$ and $p=0$, reveals $2\ell$ azimuthal intensity undulations. 
    (b) The experimental azimuthal $\ell$-mode spectra for vortex beams generated by $p$-plates with $\ell=3,5,10$. The inset shows small peaks at contributions of opposite topological charge $-\ell$, which constitute a strong visual effect in (a).
    Illustration of the formation of `ghost' OAM orders as a result of a small deviation in the grating depth, $\delta$, from $2\pi$ (dashed line) to $2\pi M$ (shaded area) in (c) an azimuthal phase-only grating (which maps to a blazed grating in polar coordinates) and (d) a phase grating with polarisation conversion (as in the $p$-plate). In both simulations, each device is designed to impart a topological charge of $\ell=5$. 
    (e) The simulated azimuthal $\ell$-mode spectra for vortex beams generated in (c) and (d) showing `ghost' OAM contributions.
    }
    \label{fig4}
\end{figure*}

\textbf{Ghost diffraction orders in azimuthal gratings.} In the experimental intensity images shown in Figure~\ref{fig3}, we observe azimuthal undulations in the intensity, reminiscent of a pearl necklace. These `pearls' are an artifact in vortex beams generated by azimuthal phase gratings and have remained unnoticed in literature, despite being present in previous works~\cite{devlin2017arbitrary,nape2020enhancing}. To understand their origin, we consider the case of a $p$-plate whose azimuthal phase is imparted purely by propagation phase in addition to a polarisation conversion to the orthogonal polarization state (see the supplemental for a discussion on $J$-plate and $q$-plate devices). The intensity distribution for a $\mbox{LG}_{5,0}$ mode generated from a $p$-plate features $2\ell$ pearls, as shown in Figure~\ref{fig4}a. Their azimuthal structure alludes to an undesirable contribution of OAM, reminiscent of the intensity distribution of a superposition of interfering vortices of opposite charges, $\ell$ and $-\ell$, which similarly form well-defined $2\ell$ petal structures. This is confirmed in the azimuthal mode decomposition performed on the vortex beams generated by the $p$-plate devices for $\ell=3,5,10$, as presented in Figure~\ref{fig4}b. Peaks can be seen at the designed azimuthal index, with up to 97\% of the power of the generated vortex being in the designed $\ell$ mode. Additional peaks of the order of a percent are seen at $-\ell$ contributions. Despite being small in contribution, the interference of these modes constitute a strong visual effect. Indeed, unlike superpositions of radial modes, whose intensity profiles scale with the radial index $p$, superpositions of opposite OAM charges lie on the same radius and have a larger overlap.

The origin of these superfluous OAM orders dates back to the work of Rowland and Michelson~\cite{rowland1893xxxix,michelson1903spectra} on false spectral lines produced by ruled gratings, in which `ghost' orders were described as originating from periodic errors of grating lines, with small error amplitudes resulting in excessively large ghost intensities. We advance the argument that these ghost orders are not limited to linear gratings, but are susceptible in any diffraction grating elements, including azimuthal gratings. In the case of the latter, the relation becomes apparent when the azimuthal profile is mapped to polar coordinates, revealing a linear blazed grating (see Figure~\ref{fig4}c). It follows then that a small deviation in the grating depth, $\delta$, which in the case of metasurfaces may result from fabrication errors that deviate from the designed pillar size or height, will lead to a cross-coupling of adjunct `ghost' OAM modes. 

To validate their presence we performed numerical simulations, in which we show that a small deviation (on the order of 10\%) in azimuthal grating depth produces the so-called `ghost' OAM orders. We consider two cases for generating a $\ell=5$ vortex beam. The first which applies a phase-only azimuthal phase (see Figure~\ref{fig4}c), and the second which applies an azimuthal phase as well as an arbitrary polarisation conversion, as in the case of $p$-plates (see Figure~\ref{fig4}d). The vortex phase of the former unwraps to a blazed grating with $\ell$ phase jumps in polar co-ordinates. In the case of the latter, the form birefringence of the metasurface structure allows each orthogonal polarization state to be addressed with an arbitrary phase profile. For $p$-plates, the azimuthal phase profile seen by the orthogonal polarization states is the same. This results in an unwrapped blazed grating with $2\ell$ phase jumps. The presence of `ghost' OAM orders, in both cases, is confirmed in the simulated azimuthal modal decomposition, shown in Figure~\ref{fig4}e. For the phase-only approach, `ghost' OAM orders appear at multiples of the designed charge $\ell$, whereas for the $p$-plate device , there is only one appreciable OAM diffraction peak at $-\ell$. The polarization conversion in the $p$-plate device results in a cross-coupling of ghost orders that suppresses their effect, with the OAM orders being spaced by $2\ell$ (see supplementary information). In both cases, the contribution at $-\ell$ manifesting as the $2\ell$ `pearls' in the intensity profile of the vortex beam. Consequently, the numerical result for the $p$-plate device agrees with the experimental results, in that a small deviation in the grating depth manifests as small contributions of `ghost' OAM orders, with the first OAM diffraction order at $-\ell$. Nonetheless, we emphasize that although this is a visually pronounced effect, it corresponds to a very small contribution of around a few percent. The azimuthal mode purity of the generated beams is above 97\%, which is comparable to the radial purity results for the same devices. 

\section{Conclusion} 
We have demonstrated the generation of high-purity vortex beams from a single metasurface, with complete control of both azimuthal and radial components. The ability to structure light vectorially and, therefore, to use polarization as a drop-port, provides a compact on-axis conversion system that is easy to implement in practical optics experiments and paves the way for its use in integrated optical devices. Furthermore, the sub-wavelength resolution of the device allows for the structuring of pure vortex beams of very high OAM, highlighted in the demonstration of a vortex beam with a $\ell=50$ charge. Our results show the benefit of using $p$-plates devices for radial mode purity, which outperform other well-established vortex-generating metasurfaces ($J$-plates and $q$-plates). As such, $p$-plates could be of interest in any vortex application in which radial mode purity cannot be neglected. 

The generalised approach is not limited to pure vortex modes, but rather allows us to harness the full resource of LG modes with any desired $\ell$ and $p$ mode distribution. In fact, the ability to perform phase and amplitude modulation, allows to sculpt many other family of paraxial beams. Exploiting the control over the complete transverse spatial degree-of-freedom is significant for the emerging field of high-dimensional quantum information, boosting communication channels with higher encoding capacities~\cite{trichili2016optical} and increasing noise robustness in entanglement distributions~\cite{valencia2021entangled}. We foresee these devices as a very convenient and powerful approach, which could further drive the uptake of higher order vortex modes.

\section{Methods} 
\textbf{Metasurface fabrication.}
The dielectric metasurface devices consist of amorphous silicon nanopillars on a silica substrate, arranged in a hexagonal closed-packed lattice and are designed to operate in the near-infrared at 1064 nm. Each pillar has a fixed height of 600 nm, while the dimensions and orientation angle of the pillars can vary to impart a different phase delay between x and y components of the field (by exploiting form birefringence). Our metasurfaces employ either geometric or propagation phase to impart the required transformations. The phase library of nanopillars was simulated using a finite-difference time-domain software (Lumerical) and a complex-value refractive index measured by ellipsometry. The fabricated devices are 200 $\mu$m in diameter and where applicable are designed to accept a focused Gaussian with beam waist of 92 $\mu$m. The metasurfaces were fabricated using electron-beam lithography (Raith 150-Two e-beam), followed by plasma enhanced reactive ion etching (Sentech SI500).

\textbf{Measuring the purity of vortex beams.} 
The azimuthal and radial purity of the generated vortex beams are measured using a well established phase flattening approach, called modal decomposition~\cite{pinnell2020modal}, with the idea of unraveling the wavefront into a Gaussian-like mode. To do this, the generated beam is imaged via a 4f telescope onto a phase-only SLM (HOLOEYE GAEA-2), on which we display complex amplitude holograms of the conjugate modes in our LG basis as outlined in Ref.~\citenum{pinnell2020modal}. We compute the optical overlap by means of a lens and measuring the Fourier-plane on-axis pixel intensity using a camera. A pinhole is placed before the camera to block unwanted diffraction orders. In this way, we extract the expansion coefficients that describe the contribution of each of the modes in our basis. We note that a combination of quarter- and half-waveplates (not shown in Figure S8) were used to ensure the correct polarisation conversion in the generation and detection steps. 

\begin{acknowledgement}

This work has been financially supported by the European Research Council (ERC) under the European Union’s Horizon 2020 research and innovation programme ``METAmorphoses”, grant agreement no. 817794. This work has been supported by Fondazione Cariplo, grant no. 2019-3923.

\end{acknowledgement}

\bibliography{bibliography}

\end{document}


\cleardoublepage

\subsection{`Ghost' OAM orders generated by $J$-plates and $q$-plates}
\begin{figure*}[!ht]
    \centering
    \includegraphics[width=1\textwidth]{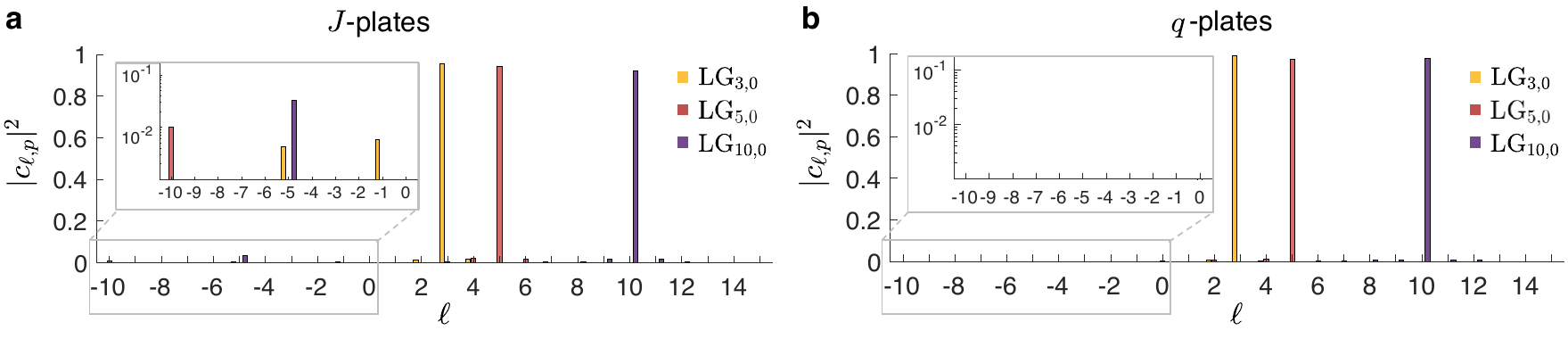}
    \caption{
    Corresponding azimuthal $\ell$-mode spectra of vortex beams generated using phase-only control in (a) $J$-plates and (b) $q$-plates, obtained via a modal decomposition in the $LG_{\ell,p}$ basis. For each type of device, three devices were designed to impart the topological charge $\ell = 3,5,10$. We note that no spatial filtering was used in the generation or detection of the vortex beams.
    }
    \label{figS1}
\end{figure*}

In Figure 4a of the main text, we presented and discussed the azimuthal $\ell$-mode purity of the vortex modes generated by the $p$-plate devices. Here, we extend those results to include the azimuthal $\ell$-mode decomposition results for the phase-only metasurface devices, namely the $J$-plates and $q$-plates, shown in Figure~\ref{figS1}. Recall that the form birefringence of metasurfaces allows to impart two different phase profiles, i.e., $\ell_1,\ell_2$ on two orthogonal polarization states, $|\lambda_+\rangle$ and $|\lambda_-\rangle$, respectively. This allowed us to design and fabricate only two $J$-plate devices that act on circularly polarized light: the first with $\ell_1=1,\ell_2=3$ and the second with $\ell_1=5,\ell_2=10$. On the other hand, the $q$-plate design is restricted to only impart conjugate OAM, $\ell_1=-\ell_2$, to orthogonal circular polarization states, such that three $q$-plate devices were fabricated, one for each charge $\ell=3,5,10$. In the case of $J$-plates, the decomposition reveals peaks at the desired charges $\ell=3,5,10$ respectively, with `ghost' OAM contributions of a few percent at the conjugate of the designed charge. For $q$-plates nearly all power is in the desired $\ell$ mode, with no auxiliary OAM contributions. To understand the origin of these OAM diffraction orders, or the lack there of, we expound the description in the main text to include the effect of a change in phase depth in azimuthal gratings for the case of $J$-plates and $q$-plates.

\begin{figure*}[h!]
    \centering
    \includegraphics[width=1\textwidth]{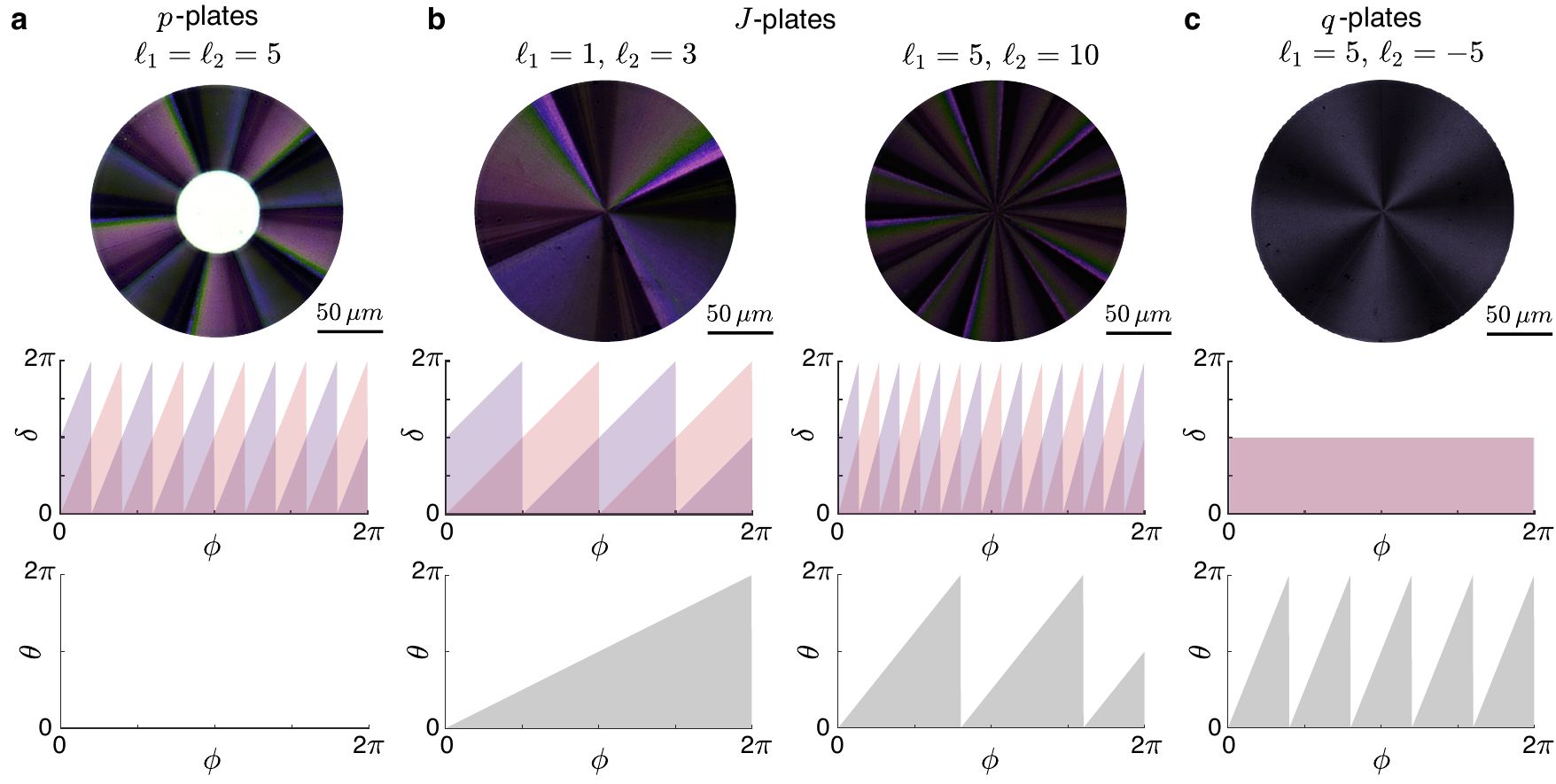}
    \caption{
    Optical images of the fabricated metasurfaces for (a) $p$-plate, (b) $J$-plates and (c) $q$-plates, with the OAM charge $\ell_1$, $\ell_2$ imparted on each orthogonal polarization state, respectively. 
    Th middle row shows the unwrapped phase delays $\delta$ for both designed orthogonal polarization states as a function of azimuthal angle $\phi$. The bottom row shows the rotation angle $\theta$ of the nano pillar for a fixed radial coordinate as a function of azimuthal angle $\phi$. 
    }
    \label{figS2}
\end{figure*}

Recall that the $p$-plate devices were designed such that $\ell_1=\ell_2=\ell$. Unwrapping the required azimuthal phase profile reveals a blazed grating with $\ell$ phase jumps for each orthogonal polarization, as seen in Figure~\ref{figS2}a, with a total of $2\ell$ phase jumps. These desired azimuthal phase delays are imparted via a propagation phase that is related to the dimensions of the nano pillar. This manifest visually as $2\ell$ wedge sectors in the device design. We note that the rotation angle for the nano pillars is kept constant in the azimuthal direction and only varies in the radial direction, so as to structure the beam vectorially and sculpt the desired amplitude profile. Since the imparted phase depends on the dimensions of the pillars, any small deviation from the designed pillar size will affect the grating depth. As discussed in the main text, such a small deviation in the grating depth leads to a cross-coupling of unwanted `ghost' OAM modes, with the first OAM diffraction order being at $-\ell$. More specifically, the `ghost' orders are formed at multiples of $|\ell_1|+|\ell_2|=2|\ell|$, centered about the desired charge. 

Next, we extend this to the generalized case of the $J$-plates, which combine both propagation phase and geometric phase control to impart orthogonal circular polarization states with different amounts of OAM, $\ell_1$ and $\ell_2$, while also performing a polarization conversion. The required phase delays are given by $\delta_+ =1/2(\ell_1+\ell_2)\phi$ and $\delta_-=1/2(\ell_1+\ell_2)\phi-\pi$, and the rotation angle of the pillars is given by $\theta=1/4(\ell_1-\ell_2)\phi$ \cite{devlin2017arbitrary}. Figure~\ref{figS2}b shows the required unwrapped phase delays, in which one can count $|\ell_1|+|\ell_2|$ phase jumps, and the corresponding rotation angles, which vary along the azimuthal coordinate, for the two fabricated $J$-plate devices. Following the same reasoning as for the $p$-plates, `ghost' OAM diffraction orders appear at multiples of $|\ell_1|+|\ell_2|$, centered about the desired charge. This is indeed the case in Figure~\ref{figS1}, with the first order contributions at the conjugate of the desired charge. An exemplary case is that of the desired charge of  $\ell_2=3$, where we see a first order peak at $\ell=-1$ and a second order peak at $\ell=-5$ which are equally spaced by $|\ell_1|+|\ell_2|=1+3=4$ modes. Similarly, for the device with $\ell_1=5$ ($\ell_2=10$), we see a first order peak at the conjugate of the orthogonal charge, $-\ell_2=-10$ ($-\ell_1=-5$).

Lastly, we turn our attention to the azimuthal decomposition of the $q$-plate devices, in which we see no `ghost' OAM orders. In contrast to the $p$-plate and $J$-plate devices, the $q$-plate device imparts the desired azimuthal phase using only geometric phase. Essentially, the device is designed using identical pillars, whose orientation angle varies in azimuth to impart the conjugate azimuthal phases $\ell_1=-\ell_2$ to orthogonal circular polarization states. The corresponding phase delay and rotation angles are shown in Figure~\ref{figS2}c. 
As opposed to the case of $p$-plates and $J$-plates, fabrication errors in the size of the pillars affect the efficiency of the polarisation conversion and not the azimuthal grating depth, as the latter is geometrically defined by the rotation angles of the pillars. As a result, the generation of `ghost' OAM orders is less susceptible in geometric phase elements, although the effect is still possible if the applied azimuthal phase grating is changed in some other way. Nevertheless, we reiterate that $q$-plates are not suitable in the context of this paper, as they do not structure the beam amplitude (they are phase-only devices) and thus generate vortex modes that unravel during propagation into a superposition with many radial modes.

\subsection{Supplementary figures}

\begin{figure}[!ht]
    \centering
    \includegraphics[width=0.65\linewidth]{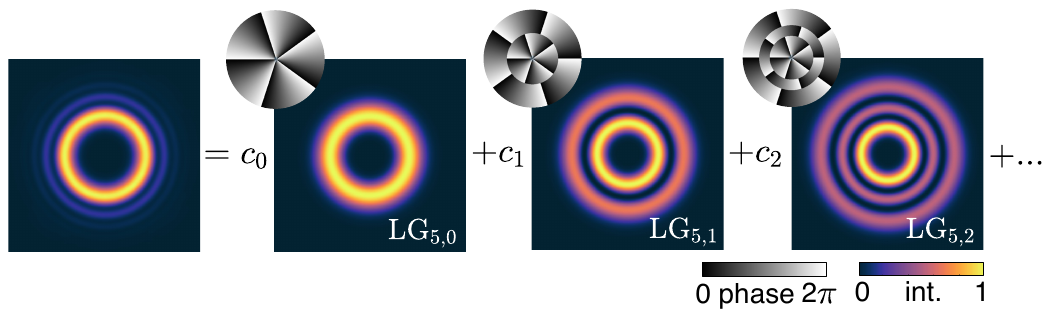}
    \caption{
    A simulation of a Gaussian beam modulated by an azimuthal phase aperture produces a vortex mode with many concentric intensity rings. During its propagation, the power of the desired azimuthal mode is spread over many higher-order radial modes.
    }
    \label{figS3}
\end{figure}

\begin{figure}[!ht]
    \centering
    \includegraphics[width=0.65\linewidth]{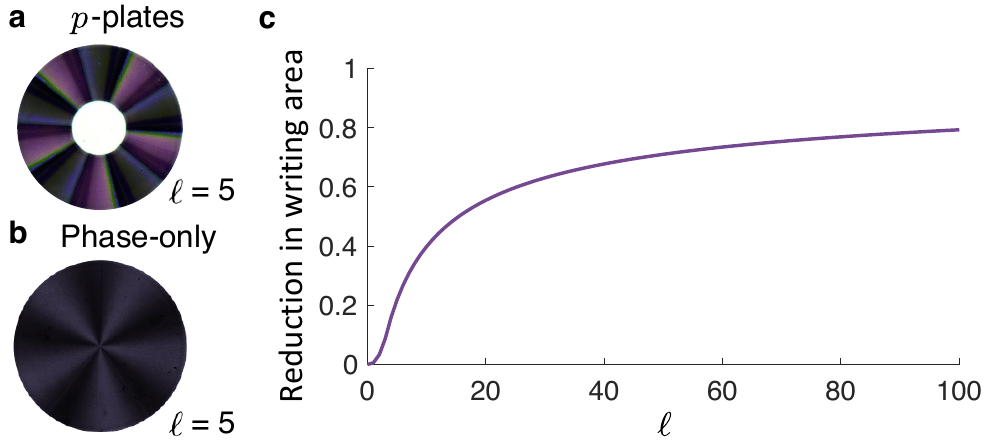}
    \caption{
    Optical image of the fabricated (a) $p$-plate and (b) phase-only metasurfaces. 
    (c) Simulation showing the reduction in the writing area of the $p$-plate device as a function of the OAM it imparts as compared to that of a phase-only metasurface. This is calculated as the complement of the area of the $p$-plate to that of the phase-only device. The active area of the $p$-plate is selected where the intensity transmission efficiency is above 5\%. This results in a characteristic ring shape of $p$-plate devices and allows to fabricate larger devices with higher OAM, while reducing the electron beam lithography writing time.
    }
    \label{figS4}
\end{figure}

\begin{figure}[!ht]
    \centering
    \includegraphics[width=0.25\textwidth]{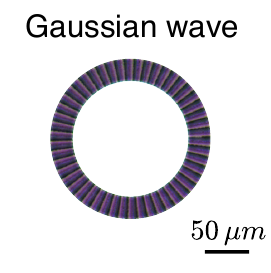}
    \caption{
    Optical image of the fabricated $p$-plate device with $\ell=50$ and $p=0$, designed to modulate an incident Gaussian beam.
    }
    \label{figS5}
\end{figure}

\begin{figure}[!ht]
    \centering
    \includegraphics[width=\textwidth]{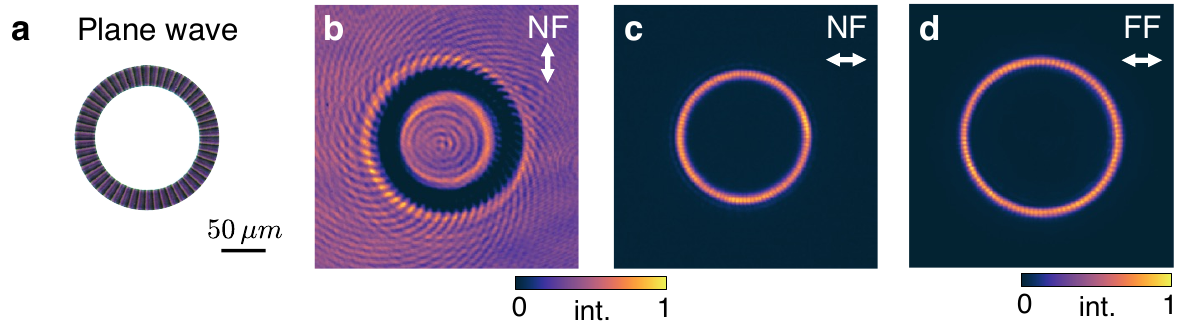}
    \caption{
    (a) Optical image of the fabricated $p$-plate device with $\ell=50$ and $p=0$, designed to modulate an incident plane wave.
    (b-c) The experimental near-field (NF) intensity distributions of the generated beam from the plane wave $p$-plate device shown in (a).
    (b) The vertically polarized component is the unwanted complementary intensity distribution from which the intensity was carved, while (c) the horizontally polarized component has the characteristic annular intensity distribution of a vortex mode. 
    (d) The corresponding far-field (FF) intensity distribution of vortex beam in (c).
    }
    \label{figS6}
\end{figure}

\begin{figure}[!ht]
    \centering
    \includegraphics[width=0.6\linewidth]{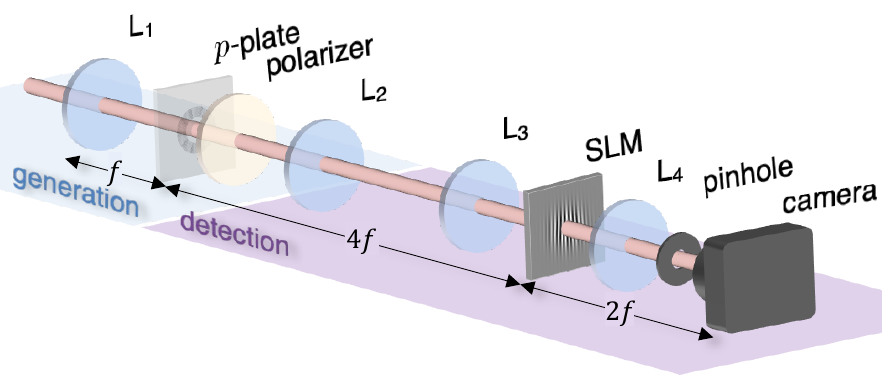}
    \caption{
    Schematic of the generation and detection of vortex beams from a $p$-plate metasurface. A focused, linearly polarized Gaussian beam propagates through the $p$-plate and exits the polarizer as a pure vortex beam. The purity of the vortex mode is characterized by performing optical overlap measurements. In the near-field (NF) of the beam, complex-amplitude holograms are displayed on a spatial light modulator (SLM) and the resulting on-axis intensity in the Fourier plane is measured using a camera. The SLM operates in reflection but for simplicity is shown in transmission.  
    }
    \label{figS7}
\end{figure}

\begin{figure}[!ht]
    \centering
    \includegraphics[width=0.6\linewidth]{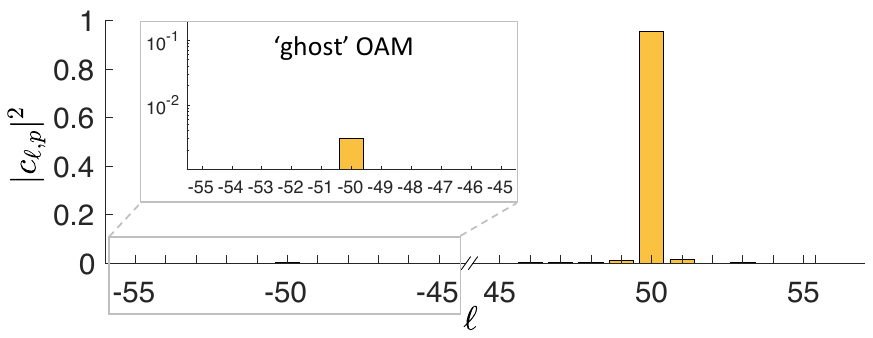}
    \caption{
    The azimuthal $\ell$-mode spectrum of the vortex beam generated by a $p$-plate with $\ell=50$ and $p=0$. The peak power contribution is at the designed charge. The inset shows a small contribution of a `ghost' OAM order at the opposite charge $-\ell$. This small contribution results in the azimuthal intensity undulations, so called `pearls', in far-field intensity image shown in the inset of Figure 2d.  
    }
    \label{figS8}
\end{figure}

\clearpage

\bibliography{bibliography}